\documentclass{PoS}

\PoS{PoS(WC2004)043}

\usepackage{epsfig}

\renewcommand{\d}{\mathrm{d}}
\newcommand{\e}{\mathrm{e}}
\newcommand{\st}{$\sigma_{\mathrm{tot}}$ }

\newcommand{\re}{\mathrm{Re\:}}
\newcommand{\im}{\mathrm{Im\:}}

\title{Derivative Dispersion Relations}

\ShortTitle{Derivative Dispersion Relations}

\author{\speaker{Regina Fonseca Ávila}\thanks{Fellow from FAPESP.}\\
        Instituto de Matem\'atica, Estat\'{\i}stica e Computa\c c\~ao 
Cient\'{\i}fica\\
Universidade Estadual de Campinas, UNICAMP\\
13083-970 Campinas, SP, Brazil\\
        E-mail: \email{rfa@ifi.unicamp.br}}

\author{M\'arcio Jos\'e Menon\\
        Instituto de F\'{\i}sica Gleb Wataghin\\
Universidade Estadual de Campinas, UNICAMP\\
13083-970 Campinas, SP, Brazil\\
        E-mail: \email{menon@ifi.unicamp.br}}

\abstract{We discuss some
analytical and numerical aspects related to the 
replacement of integral
dispersion relations by derivative  relations
and also the practical applicability of the derivative approach
in the investigation of  high-energy elastic
hadron-hadron scattering. 
Making use of a Monopole Pomeron model and singly
subtracted integral and derivative dispersion relations,
we present the results of fits to the
experimental data on the total cross sections
and the ratio of the real to the imaginary part of the forward
elastic scattering amplitude (proton-proton and antiproton-proton 
interactions). The emphasis is on the
region of low energies and, in particular, we show that once the 
subtraction constant is used as a free fit parameter
the derivative approach is equivalent to 
the integral approach even below the energy cutoff of the fitted data.}

\FullConference{Fourth International Winter Conference on Mathematical
Methods in Physics\\
09 - 13 August 2004\\
Centro Brasileiro de Pesquisas Fisicas (CBPF/MCT), Rio de Janeiro,
Brazil}

\begin{document}

\section{Introduction}

Dispersion relations constitute a traditional and important 
mathematical method in several
areas of physics, not only as a formal theoretical result, 
as a powerful phenomenological 
framework
\cite{geral}. The practical applicability of the dispersion 
relations
in high-energy particle interactions has received a renewed interest,
 mainly due to the lack
of a pure field theoretical description of the strong (hadronic) 
interactions in the sector named
\emph{soft diffractive scattering} (small momentum transfer 
and therefore large distances), which includes 
diffraction dissociation 
and elastic
scattering. In fact, despite the success of the Quantum 
Chromodynamics (QCD)
in the treatment of hard and semi-hard scattering (large and 
medium momentum transfer,
respectively), the increase of the coupling constant when going 
to the soft sector
does not allow the use of perturbative techniques and presently
 it is not know how
to obtain the soft scattering states, directly, from a pure 
nonperturbative QCD approach.

At this stage  general principles and theorems of the
 underlying field theory,
as unitarity, analyticity and crossing, constitute a 
formal framework for
phenomenological developments aimed to work as a bridge 
between  microscopic
concepts (quarks and gluons) and the experimental data.
In this context dispersion relations, as  consequence of 
the principles of analyticity
and crossing, play an important role, since they provide 
connections between the real
and imaginary parts of the elastic scattering amplitude 
as function of the energy. 
Originally they have been
introduced in the form of \emph{integral dispersion relations} 
(IDR) \cite{geral, idr}. However, despite the
important results that have been obtained, two limitations
 characterize the applicability
of the integral forms, namely their
non-local character (in order to obtain the real part of the 
amplitude
the imaginary part must be known for all values of the energy) 
and the restricted
classes of functions that allow analytical
integration. 

These shortcomings have been avoided in the region of
high energies (center-of-mass energy above 10 GeV)
with the introduction of
dispersion relations in the differential form \cite{ddr}, the
\emph{derivative dispersion relations} (DDR), which have provided
new insights in the dispersion relation techniques. 
A recent review and critical analysis on the formal and practical aspects
of the DDR, with references to outstanding results and works on the subject,
may be found in \cite{amnpa} (see also \cite{kf} and \cite{others}).

In this communication, we investigate the real 
and imaginary parts of
the elastic scattering amplitude, from proton-proton ($pp$) and
antiproton-proton ($\bar{p}p$) interactions, by means of both IDR and DDR with one
subtraction. Making use of a Monopole Pomeron model 
we obtain good descriptions of the experimental data on the total cross sections
and the ratio of the real to the imaginary part of the forward
amplitude.
Our focus is the region of low energies and we show that,
once the subtraction constant is used as a free 
fit parameter,
the derivative approach is equivalent to 
the integral approach even below the energy cutoff of the fitted data.

The manuscript is organized as follows. In Sec. 2 we recall the physical 
quantities associated with the hadronic elastic scattering to be
be investigated through the dispersion relations. In Sec. 3 we 
shortly review 
the main steps connecting IDR and DDR and in Sec. 4 we discuss the
fits to the experimental data in the context of the Monopole
Pomeron  model. The results and conclusions 
are the contents of Sec. 5.

\section{Elastic Hadron Scattering}

For equal masses elastic scattering, the complex amplitude is
expressed in terms of the Mandelstam variables, usually the
center of mass energy squared $s$ and the square of the
four-momentum transfer $t$:
$
F(s,t) = \re \ F(s,t) + \textrm{i}\ \im F(s,t)
$.
In the forward direction ($t = 0$) the most important physical quantities
that characterize the elastic scattering are the total cross section
and the $\rho$ parameter. The former is given by the optical theorem,
which at high energies reads

\begin{equation}
\sigma_{\mathrm{tot}}(s) = \frac{\im F(s,t=0)}{ s},
\end{equation}
and the latter is defined by

\begin{equation}
\rho(s) = \frac{\re F(s,t=0)}{\im F(s,t=0)}.
\end{equation}

Analyticity and Crossing connect the scattering amplitudes for
particle-particle and particle-antiparticle reactions, which are treated
as crossed channels. From the experimental point of view the $pp$ and
$\bar{p}p$ scattering correspond to the highest energy interval with
available data and for that reason our analysis will be based on these 
reactions.
In that case the above principles state that the corresponding amplitudes 
are the 
boundary value
of an analytic function of $s$ and $t$ taken as complex variables. 
The amplitudes 
for these channels are expressed in terms of crossing even (+)
and odd (-) functions of the kinematical variables:

\begin{equation}
F_{pp} = \frac{F_{+} + F_{-}}{2},
\qquad
F_{\bar{p}p} = \frac{F_{+} - F_{-}}{2}.
\end{equation}

Since \st and $\rho(s)$ are expressed in terms of the real
and imaginary parts of the scattering amplitude they constitute useful
and natural quantities to be investigate by means of dispersion
relations, expressed in terms of
crossing even and odd amplitudes, and that is the point we are interested in.

\section{Derivative Dispersion Relations}

The experimental information on the increase of the total cross sections
suggest the use of IDR with one subtraction \cite{idr}. At high energies  
and with poles removed,  they are expressed by

\begin{equation}
\re F_{+}(s)= K + \frac{2s^{2}}{\pi}P\!\!\!\int_{s_{0}}^{+\infty}
\!\!\!\d s'
\frac{1}{s'(s'^{2}-s^{2})}\im F_{+}(s'),
\end{equation}

\begin{eqnarray}
\re F_{-}(s)=  \frac{2s}{\pi}P\!\!\!\int_{s_{0}}^{+\infty} \!\!\!
\d s'
\frac{1}{(s'^{2}-s^{2})}\im F_{-}(s'),
\end{eqnarray}
where $K$ is the subtraction constant, and for $pp$ and $\bar{p}p$
scattering, $s_0=2m^2\sim 1.8$ GeV$^2$.

The detailed replacement of the above formulas by the derivative forms
may be found in \cite{amnpa}. In what follows we only recall the essential
steps and the approximations involved.
For the even amplitude we define
$s'=\e^{\xi'}$, $s=\e^{\xi}$ and
$g(\xi') = \im F_+(\e^{\xi'}) / \e^{\xi'}$ so that

\begin{displaymath}
\re F_+(\e^{\xi}) - K=
\frac{2\e^{2\xi}}{\pi}P\!\!\int_{\ln s_0}^{+\infty}
\!\!\frac{g(\xi')\e^{\xi'}}{\e^{2\xi'}-\e^{2\xi}}\d\xi'
= \frac{\e^{\xi}}{\pi}P\!\!\int_{\ln s_0}^{+\infty}
\!\!\frac{g(\xi')}{\sinh(\xi'-\xi)}\d\xi'.
\end{displaymath}

For the \emph{class of functions that are entire in the logarithm of s},
we perform the Taylor expansion of the function $g(\xi')$ and then
integrate term by term.
At the \emph{high energy limit},
we consider
the \emph{approximation\/} $s_0 = 2m^2 \rightarrow 0$,
which allows to express

\begin{displaymath}
\re F_+(\e^{\xi}) - K=
\frac{\e^{\xi}}{\pi}\sum_{n=0}^{\infty}\frac{g^{(n)}(\xi)}{n!}P
\!\!\int_{-\infty}^{+\infty}\!\!\frac{(\xi'-\xi)^{n}}{\sinh(\xi'-\xi)}
\d\xi',
\end{displaymath}
and, after some analytical steps \cite{amnpa}, we arrive at the
DDR for the even amplitude,

\begin{equation}
\frac{\re F_+(s)}{s}= \frac{K}{s} +
\tan\left[\frac{\pi}{2}
\frac{\d}{\d\ln s}\right]\frac{\im F_+(s)}{s}.
\end{equation}

With analogous procedure for the odd amplitude we obtain:

\begin{equation}
\frac{\re F_-(s)}{s}=
\tan\left[\frac{\pi}{2}
\left(1 + \frac{\mathrm{d}}{\mathrm{d}\ln s}\right)\right]
\frac{\im F_-(s)}{s}.
\end{equation}

These are the DDR for entire functions in $\ln s$ and intended for 
the high-energy region.

\section{Fits with the Monopole Pomeron Model}

In order to test the applicability of the DDR and compare the results
with those obtained with the IDR, we shall consider
as input a well known parametrization for the total cross section and investigate
the results for the $\rho$ parameter, obtained through simultaneous fits 
of both quantities to the experimental data on $pp$ and $\bar{p}p$ 
scattering. In 
the extended Monopole Pomeron model the total cross sections for these
reactions are given by

\begin{equation}
\sigma_{\mathrm{tot}}(s) = X s^{\epsilon} + Y_{+}\, s^{-\eta_{+} } +
 \tau Y_{-}\, s^{-\eta_{-}},
\end{equation}
where
the first term represents the exchange of a single Pomeron, the other
two the secondary Reggeons and $\tau = + 1$ ($- 1$) for $pp$ ($\bar{p}p$)
amplitudes. The couplings $X$, $Y_{+}$, $Y_{-}$ and the powers
$\epsilon$, $\eta_{+}$, $\eta_{-}$ (associated with the Pomeron and Reggeons 
intercepts) are free fit parameters.

The point is to analytically evaluate $\rho(s)$ in Eq. (2.2) from the
above parametrization and the procedure is as follows. From
Eq. (4.1) we determine the imaginary parts of the amplitudes $F_{pp}$ and 
$F_{\bar{p}p}$ by means of the optical theorem (2.1) and then
the imaginary parts of the
 crossing even and odd amplitudes by inverting Eq. (2.3).
These, in turn, are used as inputs in the IDR (3.1), (3.2) or in the
DDR (3.3), (3.4), leading to the determination of the real parts
of the amplitudes and, through (2.3), to $\rho(s)$ as given (2.2).

As demonstrated in \cite{amnpa}, with the above procedure, the results 
with the IDR read

\begin{eqnarray}
\rho(s)  & = & \frac{1}{\sigma_{\mathrm{tot}}(s)}
  \Big\{ \frac{K}{s}
\pm Y_-s^{-\eta_-}\cot \left(\eta_-\frac{\pi}{2}\right)
+ Xs^{\epsilon}\tan \left(\epsilon\frac{\pi}{2}\right)
-
Y_+s^{-\eta_+}\tan \left(\eta_+\frac{\pi}{2}\right) \nonumber\\
& + &
\frac{2}{\pi}
\sum_{j=0}^{\infty}
\bigg(
\pm  \frac{Y_-s{_0}^{1-\eta_-}}{s(2j+2-\eta_-)}
+ \frac{Xs{_0}^\epsilon}{2j+1+\epsilon}
+\frac{Y_+s{_0}^{-\eta_+}}{2j+1-\eta_+}
\bigg)\left(\frac{s_0}{s}\right)^{2j+1} \Big\},
\end{eqnarray}
and those obtained with the DDR are given by

\begin{eqnarray}
\rho(s)  & = & \frac{1}{\sigma_{\mathrm{tot}}(s)}
  \Big\{ \frac{K}{s}
\pm Y_-s^{-\eta_-}\cot \left(\eta_-\frac{\pi}{2}\right)
+ Xs^{\epsilon}\tan \left(\epsilon\frac{\pi}{2}\right)
- Y_+s^{-\eta_+}\tan \left(\eta_+\frac{\pi}{2}\right) \Big\},
\end{eqnarray}
where the signs $\pm$ apply for $pp$ ($+$) and $\bar{p}p$ ($-$)
scattering. We notice that the above analytical results, provided
by the DDR, are the same as those obtained with the IDR in the high-energy
limit, $s_0 = 2m^2 \rightarrow 0$.

The tests with the IDR or DDR have been made through fits of Eqs.
(4.1) - (4.2) or (4.1) - (4.3), respectively, to the experimental data on
$\sigma_{\mathrm{tot}}$ and $\rho$ from $pp$ and $\bar{p}p$ scattering.
We have used the experimental data compiled and analyzed by the
Particle Data Group \cite{pdg}, with the
statistical and systematic errors added in quadrature and energy above 
$\sqrt s_{\mathrm{min}}$ = 5 GeV (238 data points). The fits have been
performed with the program CERN Minuit and the errors in the
fit parameters correspond to an increase of the $\chi^2$ by one unit.

In order to investigate the effect of the subtraction constant
we have made fits taking $K$ = 0 or using $K$ as a free fit
parameter. Summarizing, we have four variants: IDR or DDR and in
each case $K$ = 0 or $K$ as a free fit parameter.

\section{Results and Conclusions}

The numerical results and statistical information from all the cases
analyzed are displayed in Table 1. The comparison of the fit results with
the experimental data are shown in Fig. 1 ($K$ = 0) and in Fig. 2
($K$ as a free fit parameter), where we have also included the data below 
the energy cutoff for the fit (2 < $\sqrt s $ < 5 GeV).

\begin{table}
\begin{center}
\begin{tabular}{|c|cc|cc|}
\hline
&\multicolumn{2}{|c|}{Integral Dispersion 
Relations}&\multicolumn{2}{|c|}{Derivative Disperion Relations}\\
             &       $K=0$         &       $K$ free       &       $K=0$         &  $K$ free           \\
\hline
$\epsilon$   & 0.0874 $\pm$ 0.0024 & 0.0895 $\pm$  0.0025 & 0.0814 $\pm$ 0.0021 & 0.0895 $\pm$ 0.0025  \\
$\eta_+$     & 0.399  $\pm$ 0.014  & 0.382  $\pm$  0.015  & 0.450  $\pm$ 0.012  & 0.382  $\pm$ 0.015   \\
$\eta_-$     & 0.533  $\pm$ 0.012  & 0.545  $\pm$  0.013  & 0.573  $\pm$ 0.015  & 0.545  $\pm$ 0.013  \\
$X$ (mb)     & 20.12  $\pm$ 0.51   & 19.61  $\pm$  0.53   & 21.44  $\pm$ 0.44   & 19.61  $\pm$ 0.53  \\
$Y_+$ (mb)   & 68.9   $\pm$ 2.1    & 66.4   $\pm$  2.0    & 78.0   $\pm$ 2.0    & 66.4   $\pm$ 2.0  \\
$Y_-$ (mb)   & -32.4  $\pm$ 1.8    & -33.8  $\pm$  1.9    & -38.8  $\pm$ 2.6    & -33.8  $\pm$ 1.9 \\
$K$          & 0                   & -48    $\pm$  15     & 0                   & 69  $\pm$ 18 \\
$\chi^2/DOF$ & 1.14                & 1.10                 &  1.64               & 1.10  \\
\hline
\end{tabular}
\end{center}
\caption{Results of the simultaneous fits to $\sigma_{\mathrm{tot}}$ and $\rho$
(238 data points)
with the Monopole Pomeron Model, either with the IDR or DDR 
and $K=0$ or $K$ as a free fit parameter.}
\label{tab1}
\end{table}

\begin{figure}
\epsfig{file=wc04-avilaf1a.pstex, width=.5\textwidth, height=.4\textheight}
\epsfig{file=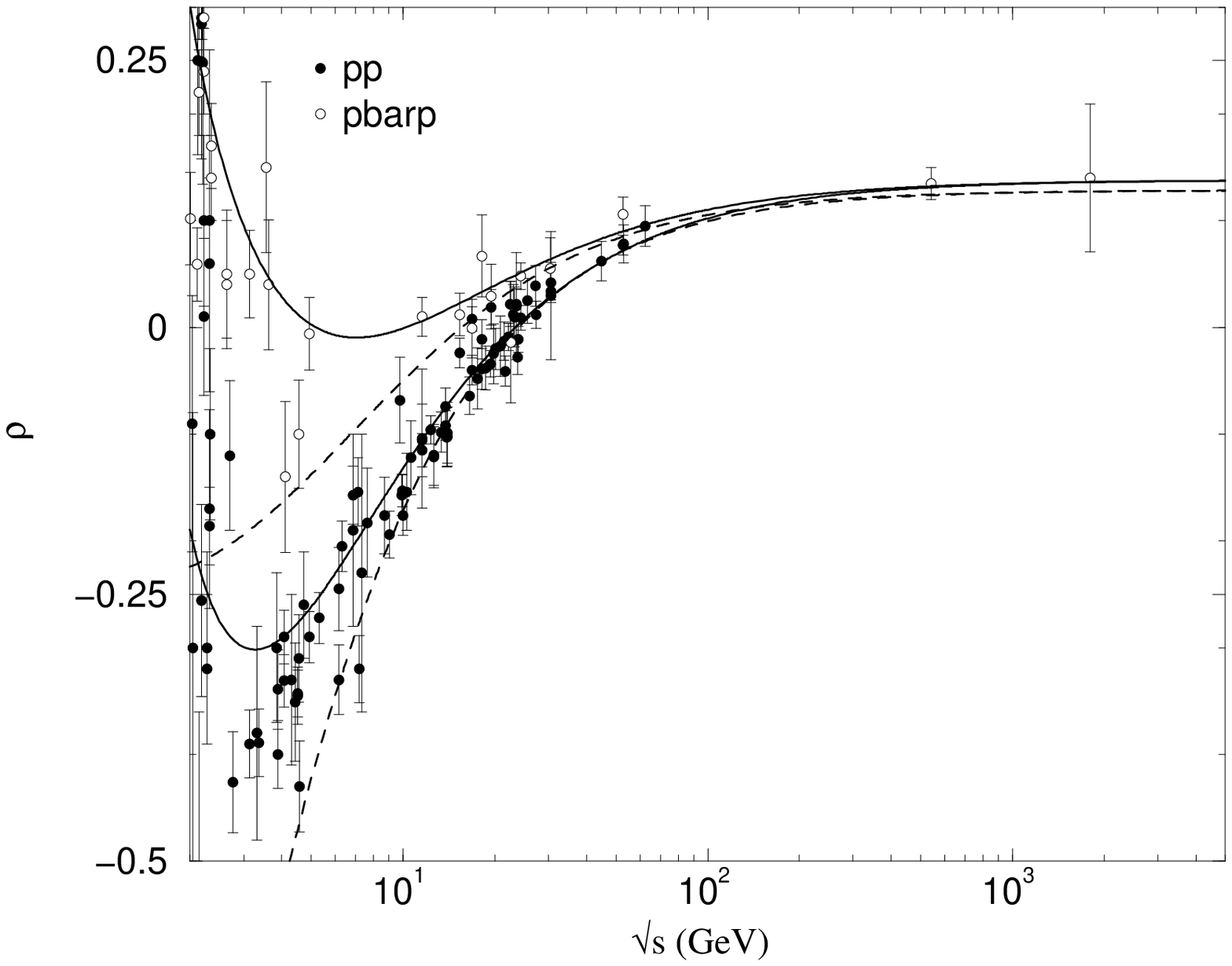, width=.5\textwidth, height=.4\textheight}
\caption{Simultaneous fit of the Monopole Pomeron model
to $\sigma_{\mathrm{tot}}$ and $\rho$ ($\sqrt s$ > 5 GeV),
\emph{assuming} $K=0$ and using either the IDR (solid) or the
DDR (dashed).} 
\label{fig1}
\end{figure}

\begin{figure}
\epsfig{file=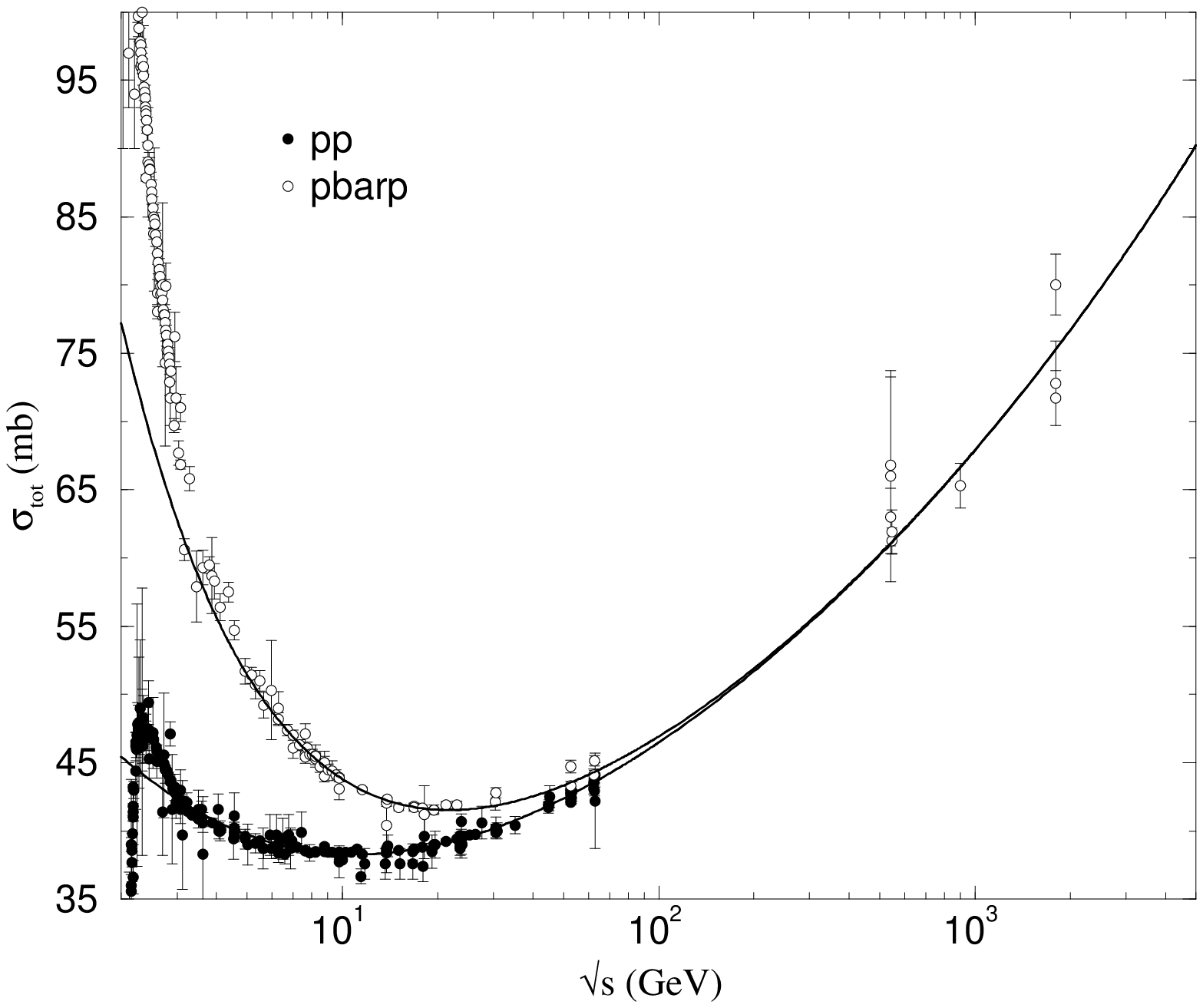, width=.5\textwidth, height=.4\textheight}
\epsfig{file=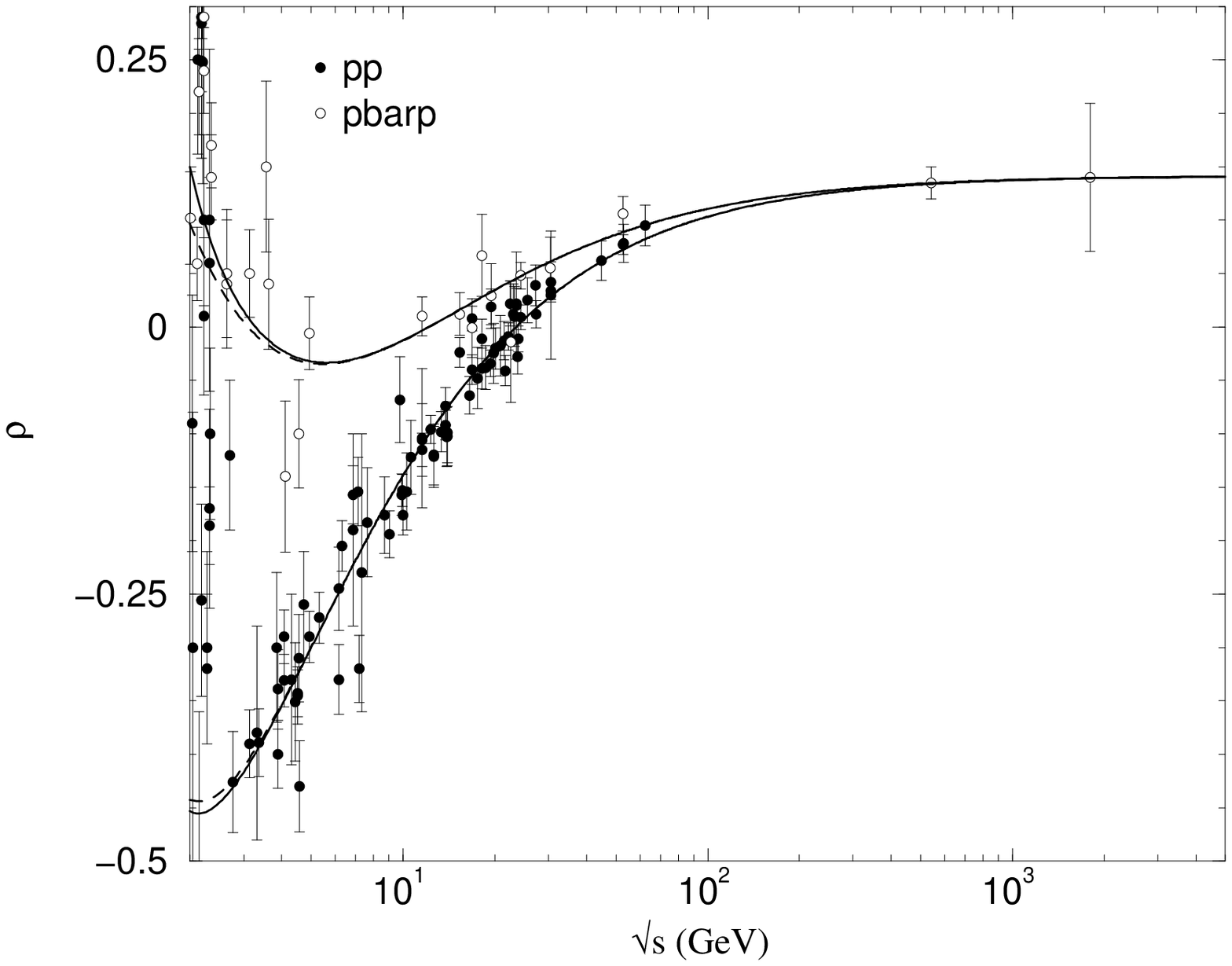, width=.5\textwidth, height=.4\textheight}
\caption{Simultaneous fit of the Monopole Pomeron model
to $\sigma_{\mathrm{tot}}$ and $\rho$ ($\sqrt s$ > 5 GeV),
with $K$ \emph{as a free fit parameter} and using either the IDR (solid) or the
DDR (dashed).} 
\label{fig2}
\end{figure}

From Table 1, the best statistical results were obtained with
$K$ as a free fit parameter ($\chi^2/DOF$ = 1.10). From Fig. 1 we see 
that for $K$ = 0, the differences between the results with IDR and
with DDR are significant at both low and high energies for $\rho(s)$
and at high energies for $\sigma_{\mathrm{tot}}(s)$. On the other hand,
from Fig. 2, in the case that $K$ is a free parameter, the results with the DDR 
and with the IDR are 
coincident. That can also be seen from Table 1, since the values of the fit 
parameters (and the $\chi^2/DOF$) are the same up to 3 significant figures.
Moreover, in this case, the description of the experimental data taking part
in the fit (above 5 GeV) is quite good (see also \cite{cudell}). 

As we have discussed, the DDR bring enclosed the high-energy approximation
($s_0 = 2m^2 \rightarrow 0$ from the IDR). Despite of this, even the data
below the energy cutoff for the fit (5 GeV) is well described in the case
that $K$ is a fit parameter. The analytical explanation for this
effect may be found in Eq. (4.2): the leading contribution in the series
expansion depends on $1/s$ and, therefore, can be absorbed by the
term with the subtraction constant, since it also depends on $1/s$.
We note that the subtraction constant is a
consequence of the polynomial bound on the scattering amplitude
and therefore has a well founded
mathematical bases.
All that shows the important role played by the
subtraction constant in the practical equivalence of the integral and
derivative approaches. We conclude that $K$ can not be disregarded or assumed
as zero in any reliable use of the derivative dispersion relations.

This work was based on the Monopole Pomeron model for the
forward scattering amplitude and singly subtracted dispersion relations.
Further tests with dipole and tripole contributions, as well as twice
subtracted dispersion relations, may bring new insights on the subject.

\end{document}